\documentclass[12pt]{article}

\input epsf

\newcommand{\bea}{\begin{eqnarray}}
\newcommand{\eea}{\end{eqnarray}}
\newcommand{\beq}{\begin{equation}}
\newcommand{\eeq}{\end{equation}}
\newcommand{\nn}{\nonumber}

\title{\vspace{-4.05cm} \rightline{\normalsize RM3-TH/99-8} \vspace{3cm}
Penguin amplitudes: charming contributions\thanks{Talk given by M.~Ciuchini
at KAON `99, June 21--26, 1999, University of Chicago, Chicago, IL, USA.}}
\date{}

\author{M.~Ciuchini$^a$, E.~Franco$^b$, G.~Martinelli$^b$ and L.~Silvestrini$^c$}

\begin{document}

\maketitle
\vspace{-1cm}
\begin{center}
$^a$INFN - Sezione di Roma III and Dipartimento di Fisica,\\
 Universit\`a di Roma Tre, Via della Vasca Navale 84, I-00146 Roma, Italy\\
$^b$INFN - Sezione di Roma and Dipartimento di Fisica,\\
 Universit\`a di Roma ``La Sapienza'', P.le A. Moro 2, I-00185 Roma, Italy\\
$^c$Technische Universit\"at M\"unchen, Physik Department,\\
 D-85784 Garching, Germany
\end{center}

\begin{abstract}
We briefly introduce the Wick-contraction parametrization of hadronic
matrix elements and discuss some applications to $B$ and $K$ physics.
\end{abstract}

In spite of the progresses in non-perturbative techniques, the computation
of hadronic matrix elements is still an open problem, particularly when
the final state contains more than one meson. In this case, methods based
on Euclidean field theory, such as QCD sum rules or lattice QCD, have serious
difficulties in computing physical amplitudes~\cite{Blok:1987sn,Maiani:1990ca}.
Besides the standard parametrization of hadronic matrix elements in terms of
$B$ parameters, it is useful for phenomenological studies to introduce a
different parametrization based on the contractions of quark fields inside the
matrix element.
In the following, we briefly discuss the Wick-contraction parametrization
introduced within the framework of non-leptonic $B$ decays
in ref.~\cite{Ciuchini:1997hb}.

\begin{figure}
\begin{center}
\epsfxsize=\textwidth
\leavevmode\epsffile{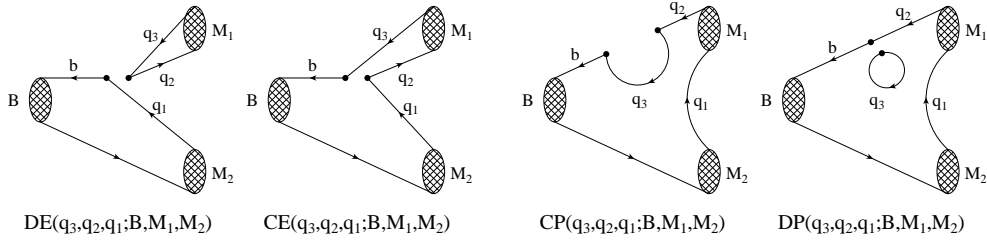}
\caption[]{\it Emission and penguin Wick-contraction topologies. The double
dots denote the operator insertion.}
\protect\label{fig:contractions}
\end{center}
\end{figure}

To be concrete, let us illustrate how this parametrization works in few
examples taken from $B$ physics.
Consider the Cabibbo-allowed decay $B^+\to {\bar D}^0\pi^+$.
Only two operators of the $\Delta B=1$ effective weak Hamiltonian
contribute to this amplitude, namely
\bea
\langle {\bar D}^0\pi^+\vert Q^{\Delta C=1}_1\vert B^+\rangle &=&
\langle {\bar D}^0\pi^+\vert \bar b \gamma_\mu (1-\gamma_5) d
~\bar u \gamma^\mu (1-\gamma_5) c \vert B^+\rangle\,,\nn\\
\langle {\bar D}^0\pi^+\vert Q^{\Delta C=1}_2\vert B^+\rangle &=&
\langle {\bar D}^0\pi^+\vert \bar b \gamma_\mu (1-\gamma_5) c
~\bar u \gamma^\mu (1-\gamma_5) d \vert B^+\rangle\,.
\eea
In this particularly simple example, the quark fields in the operators
can be contracted only according to the emission topologies $DE$ and
$CE$, shown in fig.~\ref{fig:contractions}. In the absence of a method for
computing them, these contractions can be taken as complex parameters in
phenomenological studies. The matrix elements can be rewritten as
\bea
\langle {\bar D}^0\pi^+\vert Q^{\Delta C=1}_1\vert B^+\rangle &=&
CE_{LL}(d,u,c;B^+,{\bar D}^0,\pi^+)\nn\\
&&+DE_{LL}(c,u,d;B^+,\pi^+,{\bar D}^0)\,,\nn\\
\langle {\bar D}^0\pi^+\vert Q^{\Delta C=1}_2\vert B^+\rangle &=&
CE_{LL}(c,u,d;B^+,{\bar D}^0,\pi^+)\\
&&+DE_{LL}(d,u,c;B^+,\pi^+,{\bar D}^0)\,.\nn
\eea
The subscript $LL$ refers to the Dirac structure of the inserted operators.
In general there are 14 different
topologies~\cite{Ciuchini:1997hb}--\cite{Buras:1998ra}.
Of course, in order to be predictive, one needs to introduce relations among
different parameters given by dynamical
assumptions based on flavour symmetries, chiral properties, heavy
quark expansion, $1/N$ expansion, etc.
This approach proves particularly useful for studying the $\Delta S=1$ $B$
decays. For instance, let us consider the decay $B^+\to K^+\pi^0$.
Its amplitude receives contributions from all the operators of the
$\Delta B=1$, $\Delta S=1$ effective Hamiltonian. We consider only the matrix
elements
of operators which are both proportional to the largest Wilson coefficients
$C_1$ and $C_2$ and leading order in the Cabibbo angle.
They read
\bea
\langle K^+\pi^0\vert Q^{c}_1\vert B^+\rangle &=&
\langle K^+\pi^0\vert \bar b \gamma_\mu (1-\gamma_5) s
~\bar c \gamma^\mu (1-\gamma_5) c\vert B^+\rangle \nn\\
&=& DP_{LL}(c,s,u;B^+,K^+,\pi^0)\,,\nn\\
\langle K^+\pi^0\vert Q^{c}_2\vert B^+\rangle &=&
\langle K^+\pi^0\vert \bar b \gamma_\mu (1-\gamma_5) c
~\bar c \gamma^\mu (1-\gamma_5) s\vert B^+\rangle \\
&=&CP_{LL}(c,s,u;B^+,K^+,\pi^0)\,.\nn
\eea
The penguin contractions $CP$ and $DP$ are shown in fig.~\ref{fig:contractions}.
We stress the difference between penguin operators, which we have neglected
here, and penguin contractions, which can contribute to the matrix elements
of any operator. This kind of non-perturbative contributions, called
``charming penguins'' in refs.~\cite{Ciuchini:1997hb}, could
dominate $\Delta S=1$, $\Delta C=0$ $B$ decays because other contributions are
either proportional to the small Wilson coefficients $C_{3}$--$C_{10}$ or are
doubly Cabibbo suppressed, as in the case of the factorizable emission
topologies of $Q_{1,2}^{u}$.
A detailed analysis of non-leptonic $B$ decays in this framework can
be found in refs.~\cite{Ciuchini:1997rj,Ciuchini:1998dk,Ciuchini:1998ug}.
The presence of ``charming penguin'' contributions are likely to make the
na\"{\i}ve factorization approach fail in describing this class of decays.

A different, but related parametrization of hadronic matrix elements has
been recently proposed in ref.~\cite{Buras:1998ra}. In this approach, the
parameters are the suitable combinations of Wick contractions and Wilson
coefficients which are renormalization scale and scheme independent.
In this way, the relations among contractions enforced by the renormalization
group equations are explicit. Besides, the phenomenological determination
of the parameters do not depend on the choice of the Wilson coefficients.
On the other hand, imposing relations among parameters based on
dynamical assumptions may be more involved.

Matrix-element parametrizations are less useful when applied to $K$ decays,
because there are few decay channels to fix the parameters and test the
assumptions.~\footnote{Indeed, in the case of $K\to\pi\pi$, there are only two
complex amplitudes corresponding to the $\pi\pi(I=0,2)$ final states.}
In addition, chiral relations allow the connection between matrix
elements with one pion and those with two or more pions in the final states,
the former being calculable with lattice QCD. However, a reliable lattice
determination of $\langle\pi\pi\vert Q_6\vert K\rangle$, the dominant
contribution to $\varepsilon^\prime/\varepsilon$~\cite{Buras:Kaon99}, is
presently missing~\cite{Pekurovsky:1998jd}.

Let us apply the Wick-contraction parametrization to
$K\to\pi\pi$ and verify whether there could be a connection between the
longstanding problem of the $\Delta I=1/2$ rule and a large value of
the matrix element of $Q_6$, as suggested by the recent measurement of
$\varepsilon^\prime/\varepsilon$.
In terms of the Buras-Silvestrini parameters~\cite{Buras:1998ra}, the
amplitudes $K\to\pi\pi$ with definite isospin are
\bea
\label{eq:kpipi}
  {\rm Re} A_2 &\sim& \frac{1}{3} \left( E_1 + E_2\right),\nn \\
  {\rm Re} A_0 &\sim& \left(-\frac{2}{3} E_1 +
  \frac{1}{3} E_2 - A_2 + P_1^\prime + P_3^\prime\right)\, , \\
  {\rm Im} A_0 &\sim& -\left(P_1 + P_3\right)\,,\nn
\eea
where $E_{1,2}$ are the emission parameters, $A_2$ is built with
annihilations and
\bea
  P_1  &\equiv &
  \sum_{i=2}^5 \Bigl(y_{2i-1} \langle Q_{2i-1} \rangle_{{\it CE}}
  + y_{2i} \langle Q_{2i} \rangle_{{\it DE}}\Bigr) \nn \\
  &+&\sum_{i=3}^{10} \Bigl(y_{i} \langle Q_{i} \rangle_{\it CP}
  + y_{i} \langle Q_{i} \rangle_{DP}\Bigr)
  +\sum_{i=2}^5 \Bigl(y_{2i-1} \langle Q_{2i-1} \rangle_{\it CA}
  + y_{2i} \langle Q_{2i} \rangle_{\it DA}\Bigr)\, ,\nn \\
    P_3  &\equiv &
  \sum_{i=2}^5 \Bigl(y_{2i-1} \langle Q_{2i-1} \rangle_{{\it DA}}
  + y_{2i} \langle Q_{2i} \rangle_{{\it CA}}\Bigr) \\
  &+&\sum_{i=3}^{10} \Bigl(y_{i} \langle Q_{i} \rangle_{\it CPA}
  + y_{i} \langle Q_{i} \rangle_{DPA}\Bigr)\, ,\nn
\eea
are the penguin-like parameters.
The notation $\langle Q_{i} \rangle_{{\it CE}}$ refers to the
connected emission with the insertion of the operator $Q_i$, etc.

Neglecting annihilations, as suggested by the large-$N$ counting or
by CPS+chiral symmetries~\cite{Bernard:1985wf}, we are left with four
parameters and three measured quantities.
It is unlikely that ${\rm Re} A_0$ is dominated by emissions,
since the large ratio ${\rm Re} A_0/{\rm Re} A_2$ would
require large cancellations between $E_1$ and $E_2$, see
eq.~(\ref{eq:kpipi}).
Therefore, in the most natural scenario, both ${\rm Re} A_0$ and
${\rm Im} A_0$ are dominated by penguin parameters. Notice that
$P_1$ and $P_1^\prime$ are different, so that no parametric relation
between ${\rm Re} A_0$ and ${\rm Im} A_0$ can be established.
However, the following relation holds
\beq
\label{eq:relation}
   P_1^\prime =
   z_1 \langle Q_{1} \rangle_{DP} + z_2 \langle Q_{2} \rangle_{CP}
   + P_1 (y\to z),
\eeq
where $y_i$ and $z_i$ are the Wilson coefficients of the 3-flavour effective
weak Hamiltionian.
Given this relation, we can envisage a dynamical mechanism to connect
the two parameters. Let us assume that $P_1^\prime \gg P_1 (y\to z)$
and that $\langle Q_{1,2}\rangle_{DP}$ and $\langle Q_{5,6}\rangle_{DP}$
share the same enhancement.~\footnote{We found that these two assumptions
are compatible.}
Arguments may be provided to assume that~\cite{inprep}
\beq
  f=\langle Q_{1} \rangle_{DP}\sim 1/N_c \langle Q_{2} \rangle_{CP}\sim
  -\langle Q_{5} \rangle_{DP}\sim -1/N_c \langle Q_{6} \rangle_{CP}\,.
\eeq
By using the experimental value of ${\rm Re} A_0$ and factorizing
the emission contractions, we extract $f$, from which we derive
\beq
   B_1=-9,\; B_2=7.5, \; B_5=B_6=1.5.
\eeq
It is interesting that the same mechanism
enhances the $B$ parameters entering ${\rm Re}A_0$ by
a factor of $\sim 10$ and those of ${\rm Im}A_0$ by a factor of $\sim 2$,
as required by the theoretical calculations to explain the experimental
data.

Alternatively, we could assume $P_1^\prime \sim P_1 (y\to z)$ in
eq.~(\ref{eq:relation}), namely that everything comes from the penguin
operators $Q_5$ and $Q_6$. This is the old suggestion of SVZ~\cite{Vainshtein:1977gf}.
Using perturbative coefficients, it is possible to show that this requires $B_6\sim 20$
in order to fit $({\rm Re} A_0)_{\mbox{exp}}$. Such a large value is
excluded by the measurement of $\varepsilon^\prime/\varepsilon$.

To summarize, a connection between the enhancement of ${\rm Re} A_0$
and a large value of $\varepsilon^\prime/\varepsilon$ cannot be established
without some assumption on the long-distance dynamics. We have presented
a simple example, which assume penguin-contraction dominance, that shows
the correct pattern of enhancements.
In this respect, models could give some insight, but quantitative predictions may prove
hard to produce. Hopefully, non-perturbative renormalization and
new computing techniques will help overcoming the problems which prevent the
lattice computation of ${\rm Re}A_0$ and $B_6$~\cite{Blum:Kaon99}.

M.C. and L.S. thank A.~Buras for useful discussions and excellent
steaks, beer and particularly desserts. G.M. looks forward to
acknowledging the same in the future.

\end{document}